\documentclass[twocolumn,showpacs,amsmath,superscriptaddress,aps]{revtex4-2}

\usepackage{graphicx}
\usepackage{dcolumn}
\usepackage{bm}
\usepackage{hyperref}
\usepackage{siunitx}
\usepackage{color, soul}
\usepackage{lineno}

\begin{document}

\title
{Hybrid silicon all-optical switching devices integrated with two-dimensional material}

\author{Daiki~Yamashita}
\email[Corresponding author. ]{daiki.yamashita@aist.go.jp}
\affiliation{Quantum Optoelectronics Research Team, RIKEN Center for Advanced Photonics, Saitama 351-0198, Japan}
\affiliation{Platform Photonics Research Center, National Institute of Advanced Industrial Science and Technology (AIST), Ibaraki 305-8568, Japan}
\author{Nan~Fang}
\affiliation{Nanoscale Quantum Photonics Laboratory, RIKEN Cluster for Pioneering Research, Saitama 351-0198, Japan}
\author{Shun~Fujii}
\affiliation{Quantum Optoelectronics Research Team, RIKEN Center for Advanced Photonics, Saitama 351-0198, Japan}
\affiliation{Department of Physics, Faculty of Science and Technology, Keio University, Kanagawa 223-8522, Japan}
\author{Yuichiro~K.~Kato}
\email[Corresponding author. ]{yuichiro.kato@riken.jp}
\affiliation{Quantum Optoelectronics Research Team, RIKEN Center for Advanced Photonics, Saitama 351-0198, Japan}
\affiliation{Nanoscale Quantum Photonics Laboratory, RIKEN Cluster for Pioneering Research, Saitama 351-0198, Japan}

\begin{abstract}
We propose and demonstrate hybrid all-optical switching devices that combine silicon nanocavities and two-dimensional semiconductor material. By exploiting the refractive index modulation caused by photo-induced carriers in the two-dimensional material instead of the silicon substrate, we overcome the switching performance limitation imposed by the substrate material. Air-mode photonic crystal nanobeam cavities capable of efficient interaction with two-dimensional materials are fabricated, and molybdenum ditelluride, a two-dimensional material with rapid carrier recombination, is transferred onto the cavities. The molybdenum ditelluride flake is excited by an optical pump pulse to shift the resonant wavelength of the cavity for switching operation. We have successfully achieved all-optical switching operations on the time scale of tens of picoseconds while requiring low switching energies of a few hundred femtojoules.
\end{abstract}
\maketitle

\section{Introduction}
Silicon photonics has emerged as a compelling platform for the creation of photonic integrated circuits (PICs), enabling monolithic co-integration of electronic and photonic elements on a single semiconductor chip
~\cite{Thomson:2016,Bogaerts:2018,Siew:2021}.
Within PICs, optical switching stands as an elemental and pivotal function, being responsible for routing and modulation of optical signals. While optical switches actuated by mechanical, electrical, and thermal means have been extensively studied~\cite{Soref:2018}, the increasing need for faster speed and more energy-efficient operation has shifted the focus to all-optical switching~\cite{Chai:2016}.
In particular, index modulation through photo-induced carriers is a commonly used method for controlling optical signals by another incident beam of light.

For efficient photo-excitation, all-optical switches based on microcavities are advantageous as they enhance light-matter interactions by spatially confining light
~\cite{Ibrahim:2003, Almeida:2004a, Tanabe:2005, Tanabe:2007, Hu:2008, Waldow:2008, Husko:2009, Nozaki:2010, Pelc:2014, Dong:2018, Takiguchi:2020}.
Although the use of microcavities has enabled ultrafast and energy-efficient switching, the performance is restricted by the intrinsic limitations of the silicon substrates
~\cite{Almeida:2004a, Tanabe:2005}.
To shorten the carrier lifetime of silicon for faster switching speed, ion implantation has been employed to achieve switching with a time scale of tens of picoseconds. Ion implantation, however, causes a large insertion loss for the signal light (>10~dB/cm)~\cite{Ibrahim:2003, Tanabe:2007, Waldow:2008}, which leads to inevitable performance deterioration in PICs.
While different substrate materials such as organic composite materials~\cite{Hu:2008}, III-V semiconductors~\cite{Ibrahim:2003,Husko:2009,Nozaki:2010}, and amorphous silicon~\cite{Pelc:2014}
have been utilized for higher performance, the incompatibility to fabricate these materials directly on silicon substrates remains a critical barrier, challenging monolithic on-chip integration.

One promising avenue for mitigating this challenge lies in the use of hybrid systems that integrate disparate materials onto the silicon photonics platform~\cite{Liang:2010a, Heck:2013}. Instead of direct growth, materials can be grown on separate substrates and then wafer bonded. Such hybrid systems have been effectively employed in a variety of PICs and have found utility in a range of optical devices, from lasers and modulators to photodetectors, primarily based on III-V semiconductors~\cite{Roelkens:2010,Wang:2017}.

As another candidate for this hybrid framework, two-dimensional (2D) materials with unique optoelectronic properties have attracted considerable interest
~\cite{Chen:2020,You:2020,Cheng:2021}.
Remarkably, despite their atomically thin profiles, 2D layered materials exhibit pronounced light-matter interaction even for a monolayer~\cite{Xia:2014,Huang:2022}.
Because of their diverse electronic configurations, these materials span an extraordinarily broad spectral range from ultraviolet to microwave frequencies~\cite{Chaves:2020}.
In terms of the hybridization process, 2D materials can be grown directly on various substrates over large areas using chemical vapor deposition methods, making them compatible with complementary metal-oxide-semiconductor (CMOS) technologies~\cite{Akinwande:2019}. 
In addition, the mechanical flexibility and van der Waals interfaces of 2D layered materials enable seamless integration into various photonic structures and the high uniformity at the atomic layer level ensures that light scattering loss is extremely small~\cite{Javerzac-Galy:2018}.
Owing to this optical versatility and inherent compatibility with the photonic structures, 2D material hybrids provide fruitful ground for the design and realization of next-generation all-optical switches; devices that require compact dimensions, high speed operation, superior efficiency, wide bandwidth, and cost effectiveness.

In this work, we demonstrate hybrid all-optical switching devices that combine silicon nanocavities and 2D semiconductor material. Refractive index modulation caused by photo-induced carriers in the 2D semiconductor allows us to achieve switching performance unconstrained by the properties of the silicon substrate. A thin 2D material flake characterized by rapid carrier recombination is integrated onto a silicon nanocavity. The flake placed on top of the cavity is excited by an optical pump pulse, which induces a shift in the resonant wavelength of the cavity and thereby enabling the switching operation. We have successfully implemented high speed all-optical switching with a shortened time scale of tens of picoseconds and a low switching energy of a few hundred femtojoules.

\section{Concept}

\begin{figure}[t]
\includegraphics{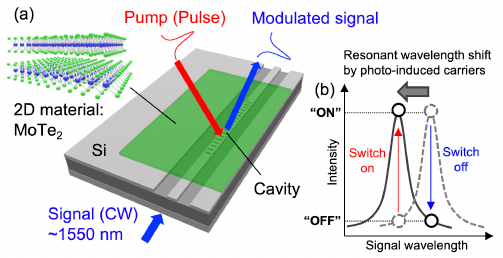}
\caption{
\label{Fig1} 
(a) Schematic diagram of our proposed hybrid all-optical switching device in which a two-dimensional material is loaded on top of a microcavity.
(b) Operating principle of all-optical switching using microcavities. Photo-induced carriers shift the resonant spectrum, enabling the selection of either switch-on or -off operation based on the signal wavelength.
}
\end{figure}

Microcavity-based all-optical switching typically relies on changes in the refractive index of the substrate material, but we propose instead to use 2D semiconductor as the refractive index changing medium. Figure~\ref{Fig1}(a) shows a schematic of the all-optical switching device using a hybrid of a silicon microcavity and a 2D material. A thin 2D semiconductor flake is loaded onto a microcavity and excited by a pump pulse to generate carriers. As shown in Fig.~\ref{Fig1}(b), the refractive index modulation by the photo-induced carriers in the flake shifts the resonant wavelength of the cavity, thereby modulating the signal light input from the waveguide.

Based on this operating principle, we first design the cavity structure for fast and low energy switching. It is necessary to reduce the device size for enhancing the interaction between light and the 2D semiconductor, and therefore we use a one-dimensional photonic crystal (PhC) nanobeam cavity~\cite{Notomi:2008,Eichenfield:2009}.
Compared to waveguide ring cavities, the PhC cavities have a larger mode overlap with low-dimensional materials
~\cite{Miura:2014,Fang:2022}.
To further enhance the interaction between light in the cavity and the 2D layered semiconductor, we employ air modes instead of the commonly used slab modes~\cite{Quan:2011,Miura:2014,Machiya:2022,Fang:2022,Yamashita:2021}.
The air modes have a large mode overlap with the 2D material loaded on top of the cavity because the electric field is distributed in the air rather than in the dielectric material, and a slight modulation of the refractive index of the 2D layered semiconductor can efficiently shift the resonant wavelength~\cite{Fang:2022}.

We now turn our attention to the 2D semiconductor used for refractive index modulation. For high speed and low energy operation, the material must be transparent in the telecom wavelength band, have large optical absorption for pump light, and have a short carrier lifetime. Molybdenum ditelluride (MoTe$_2$), which is one of the transition metal chalcogenide 2D semiconductors, meets the requirements. With a higher refractive index than silicon (Si), it allows for large modulation of the effective refractive index in the mode by photo-induced carriers.
The carrier lifetime critical for the switching time can be as fast as a few picoseconds for excitons generated by optical absorption~\cite{Chi:2019}, offering the potential for fast switching.

\section{Results and discussion}

\begin{figure}[t]
    \includegraphics{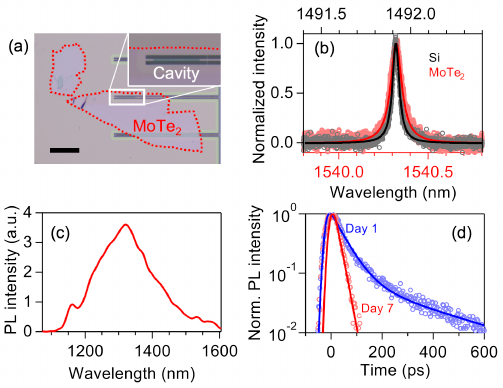}
    \caption{
    \label{Fig2} 
    (a) Optical micrograph of a hybrid Si cavity-based switching device integrated with an 11-nm-thick MoTe$_2$ flake. The scale bar is 20~$\rm{\mu}$m.
    (b) Resonant mode spectra before (black) and after (red) transfer of the MoTe$_2$ flake. The circles are data and the lines are Lorentzian fits. The resonant wavelength has shifted by 48.41~nm from 1,491.91~nm to 1,540.32~nm before and after the transfer, which is attributed to the increase in the effective refractive index of the cavity by MoTe$_2$~\cite{Fang:2022}.
    (c) PL spectrum of MoTe$_2$ on the switching device measured with continuous-wave excitation at 950~nm and 100~$\rm{\mu}$W.
    (d) PL decay curves of MoTe$_2$ on the switching device obtained with 1,000~nm and 256~fJ pulsed excitation for different dates. The circles are data, the line for Day~1 is a bi-exponential fit, and the line for Day~7 is an exponential fit. The $\tau_1$ values for Day~1 and Day~7 are 50~ps and 19~ps, respectively. The $\tau_2$ value for Day~1 is 309~ps.
    }
    \end{figure}

The hybrid switching device assembled from a PhC nanobeam cavity and a thin MoTe$_2$ flake is shown in Fig.~\ref{Fig2}(a). In implementing the cavity, we take into account the redshift after the integration of MoTe$_2$ to obtain the cavity resonant wavelength $\lambda_{\rm{cav}}$ near 1.55~$\rm{\mu}$m.
For low energy switching, a high quality factor ($Q$) of the cavity around 10,000 is desirable after transferring the 2D material. Resonant spectra are measured to investigate the change in the cavity resonant mode before and after the transfer of the MoTe$_2$ flake [Fig.~\ref{Fig2}(b)]. 
Although we observe a slight decrease in $Q$ from 33,000 to 20,000 before and after the transfer, the $Q$ remains sufficiently high for low energy switching. It also suggests that the absorption of signal light by MoTe$_2$ and additional scattering introduced by the integration are small, allowing for low insertion loss operation. 

Next, we characterize the optical properties of the transferred 2D flake to identify the band gap and to determine the carrier lifetime. Figure~\ref{Fig2}(c) shows the photoluminescence (PL) spectrum of the flake, showing that MoTe$_2$ with a thickness of $\sim$10~nm is an indirect transition semiconductor with a band gap at $\sim$1.3~$\rm{\mu}$m and the PL intensity is weak and negligible at $\lambda_{\rm{cav}}$.
The carriers dynamics is investigated by measuring the PL decay curves of the MoTe$_2$ flake on the cavity [Fig.~\ref{Fig2}(d)]. As MoTe$_2$ is known to degrade in air due to surface defect formation~\cite{Chen:2015a}, the same flake is measured on different days after the transfer. The PL lifetime for either days (Day 1: 50~ps, Day 7: 19~ps) is much shorter than Si ($\sim$ns in photonic crystals~\cite{Tanabe:2007,Fujita:2010}), thus enabling fast switching with MoTe$_2$. It is interesting to note that the increased defect density is beneficial for our devices as it boosts the switching speed, whereas faster non-radiative rates generally imply performance degradation in light-emitting devices.

\begin{figure}[t]
\includegraphics{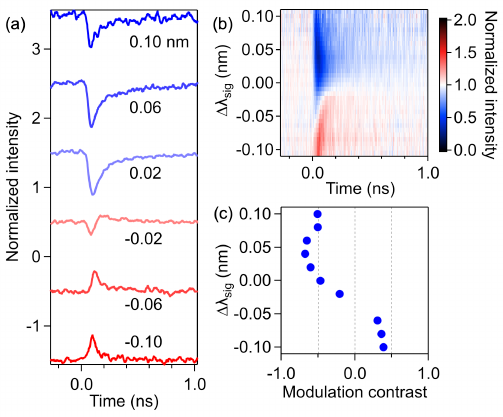}
\caption{
\label{Fig3}
(a) Switching response at different $\Delta\lambda_{\rm{sig}}$. The response is normalized by the signal intensity in the absence of a pump laser. The pump wavelength and pump pulse energy are 950~nm and 256~fJ, respectively. The curves are offset for clarity.
(b) Color plot of the switching response at different $\Delta\lambda_{\rm{sig}}$. The red color indicates peaks and the blue color indicates dips relative to the baseline of the signal intensity.
(c) Modulation contrast as a function of $\Delta\lambda_{\rm{sig}}$.
}
\end{figure}

We now examine the basic operation of the cavity-based all-optical switch.
The MoTe$_2$ flake on the cavity is pumped by a pulsed laser to modulate the cavity mode.
Figure~\ref{Fig3}(a) shows the time response of the signal light intensity with different signal laser wavelengths. Here, the signal wavelength detuning $\Delta\lambda_{\rm{sig}}$ is defined as $\lambda_{\rm{sig}} - \lambda_{\rm{cav}}$, where $\lambda_{\rm{sig}}$ is the signal wavelength.
For negative $\Delta\lambda_{\rm{sig}}$, the switching response shows a rise in the normalized intensity and then a fallback to the baseline. At $\Delta\lambda_{\rm{sig}}=-0.02~\rm{nm}$, the peak intensity decreases and a dip appears. For positive $\Delta\lambda_{\rm{sig}}$, the switching response exhibits a clear dip.
These are typical switching characteristics of cavity-based all-optical switches as shown in Fig.~\ref{Fig1}(b), where operation is switch-on for negative detuning and switch-off for positive detuning. The detailed $\Delta\lambda_{\rm{sig}}$ dependence is shown as a color plot in Fig.~\ref{Fig3}(b).

In order to achieve a large modulation in switching, we explore the optimal signal wavelength that produces the maximum modulation. We define the modulation contrast as a difference between the maximum deviation at the peak or the dip and the baseline, and present the $\Delta\lambda_{\rm{sig}}$ dependence of the modulation contrast in Fig.~\ref{Fig3}(c). Although symmetric amplitudes are expected for switch-on and switch-off operations, the resonant spectrum can become asymmetric due to free-carrier absorption~\cite{Barclay:2005, Notomi:2005, Uesugi:2006, Yamashita:2018}. The largest absolute value of the modulation contrast is obtained for a switch-off operation condition.
In the following, experiments are performed with $\lambda_{\rm{sig}}$ which maximizes the absolute value of the modulation contrast.

\begin{figure}[t]
\includegraphics{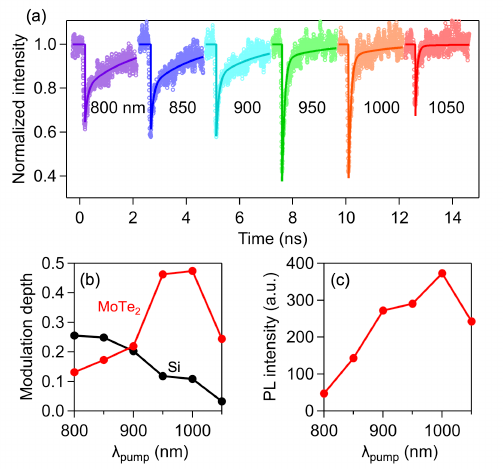}
\caption{
\label{Fig4}
(a) Switching response for different $\lambda_{\rm{pump}}$ with a pump pulse energy of 128~fJ. The circles are data and the lines are fits using Eq.~\ref{Eq1}.
(b) $\lambda_{\rm{pump}}$ dependence of modulation depth caused by the photo-induced carriers in MoTe$_2$ (red) and Si (black).
(c) $\lambda_{\rm{pump}}$ dependence of the PL intensity from the MoTe$_2$ on the cavity measured by a superconducting single-photon detector.
}
\end{figure}

For further increase of the switching performance, we look for the optimal pump wavelength recalling that the optical absorption and photo-induced carriers play a vital role in the switching dynamics. In Fig.~\ref{Fig4}(a), the switching response with different pump wavelength $\lambda_{\rm{pump}}$ is investigated.
At shorter $\lambda_{\rm{pump}}$ (800, 850, and 900~nm), the modulation depth is approximately the same and the switching response has two components in the switching recovery time.
As $\lambda_{\rm{pump}}$ increases (950~nm and 1,000~nm), the slow component in the switching recovery time decreases and the modulation becomes larger, and then the modulation decreases at $\lambda_{\rm{pump}}=1,050$~nm.
The slow component shows larger absorption at shorter wavelengths, which can be attributed to the behavior of Si. The fast component is interpreted as absorption by MoTe$_2$, with a characteristic exciton resonance near 1,000~nm~\cite{Ruppert:2014}. The observed switching recovery times are consistent with carrier lifetimes in MoTe$_2$ and Si, respectively~\cite{Tanabe:2007,Fujita:2010,Chi:2019}.

To quantitatively investigate the contributions of the MoTe$_2$ and Si components to switching, we fit the switching recovery curves. Assuming that the spectral shape of the resonant mode is Lorentzian and the shift in the resonant wavelength changes as a bi-exponential decay, the switching response can be described by~\cite{Tanabe:2007}
\begin{equation}
\label{Eq1}
    I(t)=[\{a_1e^{(-t/\tau_1)}+a_2e^{(-t/\tau_2)}\}^2+1]^{-1},
\end{equation}
where $a_1$ and $a_2$ are the amplitudes while $\tau_1$ and $\tau_2$ are the switching recovery times, with the subscripts 1 and 2 indicating the components for MoTe$_2$ and Si, respectively.
We fit the curves in Fig.~\ref{Fig4}(a) using $\tau_1=100$~ps and $\tau_2=3$~ns to obtain $a_1$ and $a_2$. In Fig.~\ref{Fig4}(b), we plot the $\lambda_{\rm{pump}}$ dependence of modulation depth caused by photo-induced carriers in MoTe$_2$ and Si. As $\lambda_{\rm{pump}}$ increases, the modulation originating from Si decreases due to a reduction in absorption.
The modulation caused by MoTe$_2$ reaches its maximum value when $\lambda_{\rm{pump}}$ is around 1,000~nm, which is consistent with the $\lambda_{\rm{pump}}$ dependence of the PL intensity shown in Fig.~\ref{Fig4}(c), indicating that more photo-induced carriers lead to larger modulation.
Consequently, to obtain large modulation and fast switching, the optimal $\lambda_{\rm{pump}}$ for this hybrid switch is $\sim$1,000~nm where optical absorption is large and the slow switching component is small.

\begin{figure}[t]
    \includegraphics{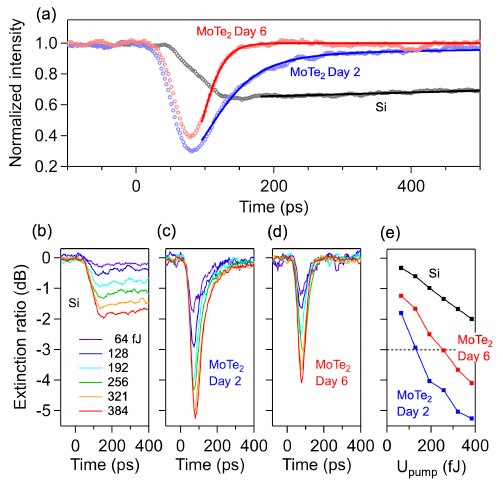}
    \caption{
    \label{Fig5}
    (a) Switching response of the Si device (black), and the MoTe$_2$ device at 2 (blue) and 6 (red) days after sample fabrication. $\lambda_{\rm{pump}}=1,000$~nm and $U_{\rm{pump}}=384$~fJ. The circles are data, and the lines are fits using Eq.~\ref{Eq1}.
    For the MoTe$_2$ Day~2 device, $\tau_1=50\pm10$~ps, $\tau_2=1,400\pm800$~ps, $a_1=0.78$, and $a_2=0.22$. Switching response of (b) Si, (c) MoTe$_2$ Day~2, and (d) MoTe$_2$ Day~6 devices at different $U_{\rm{pump}}$.
    (e) $U_{\rm{pump}}$ dependence of extinction ratio for Si, MoTe$_2$ Day~2, and MoTe$_2$ Day~6 devices.
    }
\end{figure}

Finally, with the knowledge of the optimal signal and pump wavelengths, we evaluate the switching speed and energy of this MoTe$_2$ hybrid switch.
For comparison, an all-silicon switching device with a similar $Q$ of 26,000 is also examined.
Figure~\ref{Fig5}(a) shows the switching response of the Si device and the MoTe$_2$ hybrid device, measured 2 and 6 days after sample fabrication.
The fitting is performed using a single switching time component ($\tau_1$) for Si and MoTe$_2$ Day~6 data and two switching time components ($\tau_1$ and $\tau_2$) for MoTe$_2$ Day~2 data.
The $\tau_1$ values for the Si, MoTe$_2$ Day~2, and Day~6 devices are $3,600\pm500$~ps, $50\pm10$~ps, and $33\pm5.5$~ps, respectively.
Remarkably, the switching speed of the Day~6 device is significantly faster---more than a hundredfold compared to the Si device. This acceleration is attributed to the rapid carrier recombination in MoTe$_2$, as shown in Fig.~\ref{Fig2}(d).
The lack of the slow time component ($\tau_2$) in the MoTe$_2$ Day~6 device is due to an increase in defects, leading to more absorption in MoTe$_2$ and thus less light reaching the Si substrate. The 33~ps switching speed is the fastest among PhC-cavity-based all-optical switches using Si as a substrate material~\cite{Tanabe:2005, Tanabe:2007, Dong:2018}.

Defining the extinction ratio as the logarithm of the normalized signal intensity, Figs.~\ref{Fig5}(b)-(d) display the switching response of Si, MoTe$_2$ Day~2, and MoTe$_2$ Day~6 devices with different pump pulse energy $U_{\rm{pump}}$. 
The MoTe$_2$ hybrid devices show larger modulation than the Si device, and the extinction contrast increases with increasing $U_{\rm{pump}}$ in all devices with no apparent saturation. Figure~\ref{Fig5}(e) summarizes the $U_{\rm{pump}}$ dependence of the extinction ratio value at the maximum deviation from the baseline for the three devices. The interpolated values of $U_{\rm{pump}}$  at 3~dB extinction contrast are 133~fJ and 254~fJ for the Day~2 and Day~6 devices, respectively.
These switching energies are more than one order of magnitude smaller than that of switching using Si ring cavities~\cite{Almeida:2004a, Waldow:2008} and are comparable to switching using Si PhC cavities~\cite{Tanabe:2005, Tanabe:2007, Takiguchi:2020}.

\section{Conclusion}
In conclusion, we have developed a hybrid switching device that combines a PhC nanobeam cavity and a MoTe$_2$ flake to achieve high speed and low energy all-optical operation. Despite a slight decrease in cavity $Q$ after MoTe$_2$ integration, the cavity maintains a high $Q$ essential for low energy switching. MoTe$_2$ shows faster carrier recombination compared to Si, which enables increased switching speed.
We have investigated the signal and pump wavelength dependence to maximize the modulation contrast and to improve the switching performance. Remarkably, the MoTe$_2$ hybrid device exhibits a switching time of 33~ps with a switching energy of a few hundred femtojoules, successfully overcoming the intrinsic switching speed limitation of Si while maintaining low switching energy.

To further increase the switching performance, confining the pump pulse light to the cavity resonant mode can make the switching energy low~\cite{Almeida:2004a,Tanabe:2005,Tanabe:2007,Nozaki:2010,Pelc:2014}. 
For the MoTe$_2$ flakes, fine tuning the surface defect density and thickness can improve the switching speed and switching energy. The defect density in MoTe$_2$ progressively rises with air exposure, and encapsulating it with hexagonal boron nitride should prevent further introduction of defects for stabilizing the device performance at the optimum condition~\cite{Ahn:2016}. In addition, all-optical switching operation at different pump wavelengths is possible by changing the 2D material. Our proposed structure is compatible with CMOS technologies, thus demonstrating the potential for high speed and low energy operation in integrated all-optical switching systems.

\section*{Methods}
\paragraph*{Sample preparation.}
Air-mode nanobeam cavities, with an 800~nm width and a lattice constant of $a=350$~nm~\cite{Fang:2022,Yamashita:2021}, feature a parabolically modulated lattice over 16 periods to create a central minimum optical potential. The cavities have air holes of $0.35a\times500$~nm, and the cavity center has a period of $1.18a$. The nanobeam cavities are fabricated using silicon-on-insulator wafers with a top Si thickness of 260~nm and a buried oxide thickness of 1~$\mu$m. The nanobeams are patterned by electron beam lithography, followed by inductively-coupled plasma etching of the top Si layer. The buried oxide layer under the nanobeam structures is then etched by hydrofluoric acid. MoTe$_2$ crystals are purchased from HQ Graphene. MoTe$_2$ flakes are prepared on polydimethylsiloxane (PDMS) by employing a mechanical exfoliation technique. The MoTe$_2$ flakes are transferred onto the cavities using a conventional PDMS stamping method~\cite{Fang:2022}.

\paragraph*{Cavity-resonant spectrum measurement.}
A homebuilt confocal microscopy system measures the resonant spectrum of the cavities~\cite{Yamashita:2018}. The light from a telecom-band wavelength-tunable laser is directed one end of the waveguide through a side objective lens with a numerical aperture (NA) of 0.40 and a working distance of 20~mm, and the cavity-coupled light is detected by an InGaAs photodetector through a top objective lens with an NA of 0.85 and a working distance of 1.48~mm.

\paragraph*{PL measurement.}
PL measurements are conducted at room temperature in a dry nitrogen atmosphere~\cite{Ishii:2019}. A variable wavelength Ti:sapphire laser is used as the continuous-wave light source, with its power controlled via neutral density filters. The laser beam is focused onto the samples with the top objective lens. The PL spectrum is measured through the same objective lens and detected using a liquid-nitrogen-cooled {InGaAs} diode array attached to a spectrometer.
For PL decay curve measurements, we use a picosecond supercontinuum pulsed laser (30~ps pulse width and 80~MHz repetition rate). After passing the laser through a monochromator to narrow its spectral line width to $\sim$10~nm, we directed it onto the sample using the top objective lens. The PL from the MoTe$_2$ is collected through the top objective lens and detected with a superconducting single-photon detector. A time-correlated single-photon counting module is used to collect the data.

\paragraph*{Switching measurement.}
The MoTe$_2$ flake on the cavity is pumped by the picosecond supercontinuum pulsed laser to modulate the cavity mode. The signal light is injected into one end of the waveguide through the side objective lens, and the modulated signal light is detected by a superconducting single-photon detector through the top objective lens and a bandpass filter with a transmission bandwidth of 12~nm.

\section*{Acknowledgments}
This work is supported in part by JSPS (KAKENHI JP22K14624, JP22K14625, JP20H02558, JP23H00262) and MEXT (ARIM JPMXP1223UT1141). N.F. is supported by RIKEN Special Postdoctoral Researcher Program. The FDTD calculations are performed using the HOKUSAI BigWaterfall supercomputer at RIKEN. We acknowledge the Advanced Manufacturing Support Team at RIKEN for technical assistance.

\section*{Author contributions}
D.Y. and Y.K.K. conceived and designed the experiments. D.Y. performed the experiments and analyzed the data. N.F. assisted the materials transfer, and S.F. aided the construction of measurement setup. D.Y. and Y.K.K. wrote the manuscript with inputs from all authors. Y.K.K. supervised the project.

\end{document}